\documentclass[aps,twocolumn]{revtex4}
\usepackage[dvips]{epsfig}
\usepackage{amsmath,amssymb}
\newcommand{\eq}[1]{(\ref{#1})}
\newcommand{\fig}[1]{Fig.\ref{#1}}
\newcommand{\be}{\begin{equation}}
\newcommand{\ee}{\end{equation}}

\begin{document}

\title{How long does it take to pull an ideal polymer into a small hole?}
\author{A.Yu. Grosberg$^1$\footnote{Permanent address: Department of Physics,
University of Minnesota, 116 Church Street SE, Minneapolis, MN 55455, USA}, S.
Nechaev$^1$\footnote{On leave of absence: L.D. Landau Institute for Theoretical
Physics, 117334, Moscow, Russia}, M. Tamm$^1$\footnote{Permanent address: Physics
Department, Moscow State University 119992, Moscow, Russia}, O. Vasilyev
$^{2\dagger}$ } \affiliation{$^1$Laboratoire de Physique Th\'eorique et
Mod\`eles Statistiques, Universit\'e Paris Sud, 91405 Orsay Cedex, France \\
$^2$Centre de Recherche en Mod\'elisation Mol\'eculaire, Universit\'e de
Mons-Hainaut,  Av. Copernic, 1 B-7000 Mons, Belgium}

\date{\today}

\begin{abstract}
We present scaling estimates for characteristic times $\tau_{\rm
lin}$ and $\tau_{\rm br}$ of pulling ideal linear and randomly
branched polymers of $N$ monomers into a small hole by a force
$f$. We show that the absorbtion process develops as sequential
straightening of folds of the initial polymer configuration. By
estimating the typical size of the fold involved into the motion,
we arrive at the following predictions: $\tau_{\rm lin}(N) \sim
N^{3/2}/f$ and $\tau_{\rm br}(N) \sim N^{5/4}/f$, and we also
confirm them by the molecular dynamics experiment.


\end{abstract}

\maketitle


There are models in physics - most famously exemplified by an
Ising model - whose importance is not due to their practical
applicability to anything in particular, but because they provide
the much needed training ground for our intuition and for the
development of our theoretical methods.  In polymer physics, one
of the very few such models is that of ideal polymer absorbtion
into a point-like potential well.  For a homopolymer in
equilibrium, this is a clean yet interesting example of a second
order phase transition \cite{homopolymer,Eisenregler}. For a
heteropolymer, even in equilibrium, the problem is still not
completely understood -- see, for example,
\cite{gr_sh,nech_na,madr,step}; mathematically it is similar to
the localization (pinning) transition in the solid-on-solid models
with quenched impurities \cite{forg,vannim,alex}.

Surprisingly enough, the dynamics of a polymer chain pulled into a potential well is
almost not discussed at all.  On the first glance, one might think that this
dynamics should be related to the dynamics of coil-globule collapse
\cite{DeGennes1,DeGennes2,klushin,obukhov,gold}, but the similarity between these
two problems is limited.  Indeed, globule formation after an abrupt solvent quench
begins by a simultaneous chain condensation on many independent nucleation centers,
while the absorbtion process can be viewed as pulling of a polymer with a constant
force into a single hole. Hence, only a single attractive center causes the chain
condensation. In this sense, dynamics of absorbtion into a potential well is more
similar to a driven polymer translocation through a membrane channel
\cite{Kasianowicz,
Meller3,
Storm,kar1,kar2}.

To describe in words the polymer pulled into a hole, let us
imagine a long rope randomly dropped at the desk top. Pull now the
rope by its end down from the desk edge. It is obvious, and can be
established through an experiment accessible even to theorists,
that the rope does not move all at once.  What happens is only a
part of the rope moves at the beginning, such that a fold closest
to the end straightens out. As soon as the first fold is
straightened, it then transmits the force to the next fold, which
starts moving, and so on.  The rope sequentially straightens its
folds, such that at any moment only part of the rope about the
size of a current fold is involved in the motion, while the rest
of the rope remains immobile.

In the present letter we show how this mechanism applies to an
ideal homopolymer dynamics and how it sheds light on the physics
behind scaling estimates \cite{kar2} for the relaxation times for
both linear, $\tau_{\rm lin}$, and randomly branched, $\tau_{\rm
br}$, polymers.


We address the problem on the simplest level of an ideal Rouse homopolymer of $N$
monomers, each of size $a$, embedded into an immobile solvent with the viscosity
$\eta$.  One end of the chain is put in the potential well (the "hole"), and the
polymer is pulled down to the hole with constant force $f$, acting locally on just
one single monomer, the one currently entering the hole. We will be interested in
the relaxation time $\tau$ of the complete pulling down of the entire polymer into
the hole, we want to know how $\tau$ depends on polymer length $N$ and the pulling
force $f$. Neglecting inertia, the balance of forces reads $f=f_{\rm fr}(t)$, where
friction force $f_{\rm fr}$ should be linear in velocity $\frac{dn}{dt}$, with
$n(t)$ being the number of monomers `swallowed by the hole' by the time $t$. More
subtly, friction coefficient must be proportional to the number of monomers $m(t)$
currently involved in the forced motion: $f_{\rm fr}=\eta\, a^2 \, m(t)
\frac{dn(t)}{dt}$. Our main task will be to estimate the length $m(t)$ involved in
the motion for either linear or branched polymers.

In order to clarify our approach, let us compute the
characteristic time $\tau$ of pulling down of an initially
stretched linear polymer.  In this case, all monomers move except
those already absorbed in the hole, so $m(t) = N-n(t)$, and so the
force balance condition reads
\be f = \eta\,a^2\, \left[N-n(t) \right] \frac{dn(t)}{dt} \ .
\label{eq:1} \ee
Upon integration, this yields
\be \tau_{\rm straight} = \frac{\eta\, a^2}{2f}\, N^2 \ .
\label{eq:2} \ee
In terms of dependence on $N$, this result is similar to the Rouse
relaxation time of a linear polymer, $\tau_R \simeq \tau_0 N^2$,
where $\tau_0 \simeq \eta a^3 / k_B T$, and $T$ is temperature.
Our estimate, \eq{eq:2}, remains valid as long as $\tau_{\rm
straight} < \tau_{R}$, because in this case the initially straight
polymer has no time to coil up on itself while being pulled up by
the force $f$.  Given the expression for $\tau_{\rm straight}$,
(\ref{eq:2}), the condition $\tau_{\rm straight} < \tau_{R}$
translates into $f > k_B T /a$, the latter making an obvious
sense: pulling force should be strong enough to cause significant
stretching of every bond, or, in other words, Pincus blobs must be
as small as of the order of one monomer \cite{Pincus_blobs}.


Now we turn to more realistic dynamics starting from a Gaussian
coil configuration, and look at $\tau_{\rm lin}(N)$. To begin
with, let us consider a simple scaling estimate \cite{kar2}. First
of all, we note that the only time scale relevant for the problem
is the above mentioned Rouse time $\tau_R$, which is the time
needed for the coil to diffuse over the distance about its own
size, $R \sim a N^{1/2}$. With $\tau_R$ being the solitary
relevant time scale, the absorbtion time should be written in the
scaling form
\be \tau_{\rm lin} = \tau_R \phi \left(\frac{fR}{k_BT} \right) \ ,
\label{eq:gr_1} \ee
where the factor $\phi$ can depend on all other relevant
parameters only in the dimensionless combination $fR/(k_BT)$.  The
second step follows from the fact that the speed of the process,
$N/\tau_{\rm lin}$, must be linear in the applied force, which
means that the function $\phi(x)$ should be inversely proportional
to its argument, $\phi(x) \sim 1/x$. This then yields
\be \tau_{\rm lin} \sim \tau_R \frac{k_BT}{fR} \sim \frac{\eta
a^2}{f} N^{3/2} \ . \label{eq:gr_2} \ee

\begin{figure}[ht]
\epsfig{file=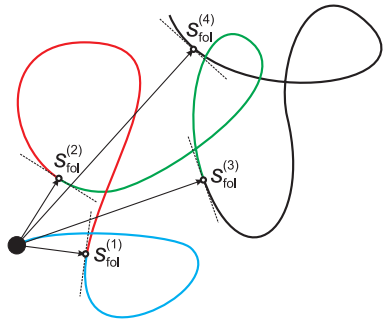, width=5cm} \caption{(Color online)
Random coil viewed as a sequence of folds.} \label{fig:1}
\end{figure}

Scaling argument gives the answer, but does not give much insight. To gain proper
insight, let us show how this same answer results from the process of sequential
fold straightening.  The definition of ``folds'' and of their lengths $s_{\rm
fol}(N)$ is illustrated in Fig. \ref{fig:1}.   Since the chain is being pulled all
the time into an immobile point in the center, the definition of a fold arises from
considering only the radial component of the random walk, $\left\vert {\mathbf r}
(s) \right|$, where ${\mathbf r}(s)$ stands for the (natural) parametric
representation of a path.  In $3D$, for example, $\left\vert {\mathbf r} (s)
\right|$ represents Bessel random process \cite{bessel1,bessel2}. To obtain lengths
of folds, $s_{\rm fol}^{(1)},\, s_{\rm fol}^{(2)}, \ldots,\, s_{\rm fol}^{(k)}$, one
has to coarse grain the trajectory up to the one monomer scale $a$, thus making a
Wiener sausage, and then find the local minima of $\left\vert {\mathbf r} (s)
\right|$ separated by the biggest maxima.  It is obvious that there are about
$\sqrt{N}$ of such minima, separated by the intervals about $\sqrt{N}$ each.

When we pull down the Gaussian chain by its end with a constant force $f$, only the
current fold, of the length $ s_{\rm fol}^{(i)} \sim \sqrt{N}$, is involved in the
forced motion and experiences friction. The equation \eq{eq:1} is still valid, but
$N$ should be replaced by $s_{\rm fol}^{(i)}$ and $n$ should be integrated in the
limits $\left[0,s_{\rm fol}^{(i)} \right]$, yielding relaxation time for one fold
\be \tau_{\rm fol}=\frac{\eta\, a^2}{2f}\,  \left(s_{\rm fol}^{(i)}
\right)^2 \sim \frac{\eta\,a^2}{f} N \ . \label{eq:8} \ee
All $\sim \sqrt{N}$ folds relax sequentially, meaning that $\tau_{\rm lin} \sim
\sqrt{N} \tau_{\rm fol}$, and thus returning the scaling answer (\ref{eq:gr_2}).

Notice that the absorbtion time of a linear chain, $\tau_{\rm lin}$, Eq.
(\ref{eq:gr_2}), is much shorter than the typical Rouse time under the very weak
condition $f > k_BT/(a\sqrt{N})$, which means Pincus blob dictated by the force $f$
should only be smaller than the entire coil.  Under this condition, during the
absorbtion process the configuration of the chain parts not involved into the
straightening of a current fold, can be considered as quenched.


\begin{figure}[ht]
\epsfig{file=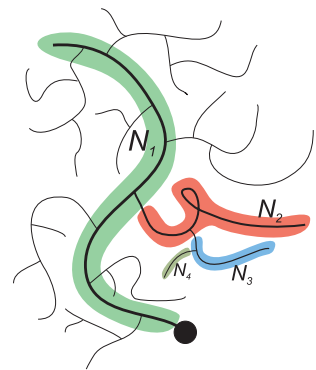,width=4cm} \caption{(Color online) Randomly branched
chain and the corresponding hierarchy of scales.} \label{fig:2}
\end{figure}

To understand the chain pulling dynamics even better, it is useful
to consider another fundamental fractal model, that of randomly
branched polymer, schematically shown in \fig{fig:2}.  To find
scaling approach for the randomly branched polymer, we have to
estimate the analog of Rouse time as the time for a polymer to
diffuse to its own size. Since all monomers experience friction
independently of each other (there are neither solvent dynamics,
nor hydrodynamic interactions), it follows that the diffusion
coefficient of a Rouse coil, linear or branched, is given by $D
\sim k_B T/(\eta a N)$. However, the radius of a branched polymer,
unlike a linear one, is only $R \sim a N^{1/4}$. Therefore, the
Rouse relaxation time for the branched polymer reads $\tau_R \sim
\tau_0 N^{3/2}$. The rest of the argument repeats exactly what we
have done for the linear chain, yielding the result
\be \tau_{\rm br} \sim \tau_R \frac{k_BT}{fR} \sim
\frac{\eta\,a^2}{f} N^{5/4} \label{eq:gr_3} \ee

Let us show now how one can re-derive \eq{eq:gr_3} generalizing the concept of
Bessel process to randomly branched chains. It is convenient to introduce a
hierarchy of scales in the $N$--link ideal randomly branched chain. Let us denote by
$N_1$ a number of monomers in a typical "primitive" (or "bare") path of a largest
scale. The size $R$ of an ideal randomly branched polymer is $R\sim a N^{1/4}$. The
bare path of the maximal scale is a Gaussian random walk with $R\sim a N_1^{1/2}$.
Hence, $N_1\sim N^{1/2}$. The average number of monomers in all side branches
$N_2,\, N_3,\, N_4$, etc. attached to {\it one} bond of the bare path of the maximal
scale (as shown in \fig{fig:2}a), is of order of $N/N_1\sim N^{1/2}$. All these side
branches form a randomly branched configuration of size $R_2\sim a (N/N_1)^{1/4}\sim
a N^{1/8}$. Now, the number of monomers in the bare path of the second scale, $N_2$,
can be obtained repeating the above scaling arguments: $R_2\sim a N_2^{1/2}$, giving
$N_2\sim N^{1/4}$, and so on: $N_3\sim N^{1/8}, \ldots, N_k\sim N^{2^{-k}}$,\ldots
-- see \fig{fig:2}.

On all scales the bare paths $N_1, N_2,\ldots N_k$, etc. form the
Gaussian sub-coils, each of which can be divided in folds $s_{\rm
fol}^{(1)},\, s_{\rm fol}^{(2)},\ldots  s_{\rm fol}^{(k)} ,\ldots$
in the same manner as it is done above for a linear chain.  For
$s_{\rm fol}^{(k)}$ we have the estimate
\be s_{\rm fol}^{(k)} \sim a N_k^{1/2}\sim a N^{2^{-(k+1)}} \qquad
(k=1,2,\ldots) \ . \label{eq:10} \ee

If the randomly branched polymer is pulled by its end with a
constant force, $f$, then all Gaussian sub-coils on all scales
straighten their folds simultaneously. The total typical number of
monomers $ s_{\rm all} $ in all folds simultaneously involved into
the motion on all scales can be estimated as
\be
s_{\rm all} \sim  s_{\rm fol}^{(1)} s_{\rm fol}^{(2)} s_{\rm fol}^{(3)}... =  a
N^{1/4+1/8+1/16+\ldots}=a N^{1/2} \ . \label{eq:11}
\ee
We have seen on the example of the linear chain, that the
characteristic time to straighten the fold of a typical length $s$
is of order of $\tau \sim s^2$ -- see \eq{eq:8}. So, as it follows
from \eq{eq:11}, the characteristic time of straightening out all
folds of length $ s_{\rm all} $ is of the order of
\be \tau_{\rm fol}^{\rm br} \sim \frac{\eta\,a^2}{2f}  s_{\rm
all}^2 \sim \frac{\eta\,a^2}{f} N \label{eq:12} \ee
In a randomly branched polymer there are $N_1 a / s_{\rm fol}^{(1)} $ independent
parts, which relax sequentially. Hence, the total relaxation time can be estimated
as $\tau_{\rm br} \sim \left(\left. N_1 a  \right/ s_{\rm fol}^{(1)} \right)
\tau_{\rm fol}^{\rm br}$, which returns the result \eq{eq:gr_3}.

\begin{figure}[ht]
\epsfig{file=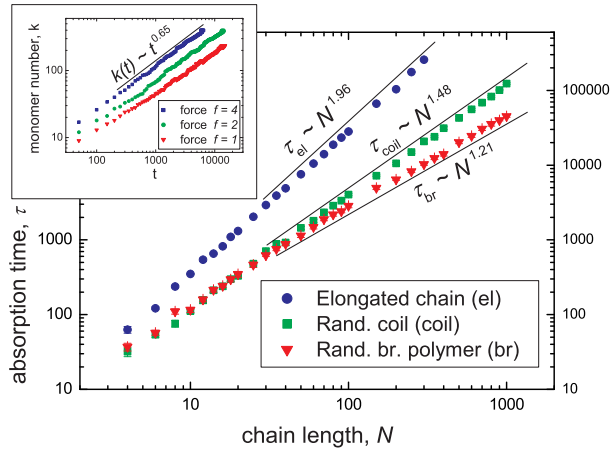,width=9cm} \caption{(Color online). Log-log plot of
relaxation time against $N$. The units: $a=1$, $k_BT=1$, and $\eta =10$. Bullets
($\bullet$) -- initially straightened linear chain, squares ($\blacksquare$) --
initially coiled linear polymer, triangles ($\blacktriangle$) -- randomly branched
polymers. The slopes  are consistent with theoretical predictions.  Inset: time
dependence of the $k(t)$ in the log-log scale.  Here $k(t)$ is the monomer number
such that at the given time moment $t$ monomers before $k$ are already strongly
involved in the biased motion towards the absorbing hole, while monomers after $k$
are still diffusing not influenced by the force.} \label{fig:Molecular_dynamics}
\end{figure}

We have tested our results by a molecular dynamics experiment. We used standard
Rouse model with harmonic bonds between neighboring monomers along the chain and
with the averaged bond length, $a=1$. The attracting force $f$ acts at any moment on
one monomer only and always points to the origin. Once a given monomer $k$ gets
inside the absorbing hole ($|{\bf r}_k|<R_{a}=1$, where ${\bf r}_k$ is the position
vector of the $k$th monomer) this monomer is considered absorbed, it does not move
any more and does not exert any force on monomer $k+1$, its neighbor along the chain
-- but instead neighbor becomes the subject of the steady pulling force $f$. The
same mechanism was simulated for the randomly branched polymer, except different
branches can be pulled in parallel. The solution of dynamical equations is realized
in the frameworks of the numerical velocity Verlet scheme \cite{verlet}. The initial
state of the linear chain was prepared by placing the first monomer at $|{\bf
r}_1|=R_{a}$. The initial configuration of the randomly branched polymer was
prepared as described in the work \cite{KCh}, and then polymer was shifted as a
whole to place one randomly chosen monomer on the absorbing surface $|{\bf
r}|=R_{a}=1$. By performing 100 runs, we have computed averaged pulling times of
linear chains from initially elongated and initially Gaussian conformations, as well
as for a randomly branched polymer. The results are shown in the Fig.
\ref{fig:Molecular_dynamics} and demonstrate very satisfactory agreement with the
theoretical predictions (\ref{eq:gr_2}) and (\ref{eq:gr_3}).

In order to check more directly our proposed mechanism of absorption, we have looked
at how Gaussian polymer chain gets involved into the biased motion driven by the
pulling force $f$. The result is shown in the inset of figure
\ref{fig:Molecular_dynamics}. In this inset, $k(t)$, marks the crossover between
current fold and the rest of the chain. The current fold is operationally defined to
include all monomers, with numbers $j< k(t)$, whose motion is currently strongly
affected by the pulling force $f$. Other monomers with the numbers $j>k(t)$ are not
yet affected. The simulation indicates that thus defined $k(t)$ follows quite
accurately the power law $k(t)\sim t^{2/3}$ with the exponent independent of the
applied force. Inverting $k(t)$, we get $t \sim k^{3/2}$, which suggests that the
time of complete absorbtion of a $k$-length subchain, which is given by our
theoretical prediction (\ref{eq:gr_2}), and the time of involving $k$-length
subchain in the force-biased motion scale in the same way.

To conclude, we have considered scaling estimates of the relaxation time associated
with pulling linear or randomly branched polymer chain into a hole by a constant
force $f$.  We found that these estimates, $\tau_{\rm lin}\sim (\eta\,a^2/f)
N^{3/2}$ and $\tau_{\rm br}\sim (\eta\,a^2/f) N^{5/4}$, are consistent with the
molecular dynamics simulation and they can be understood in terms of the scenario of
sequential straightening of the polymer parts which we called folds and which were
defined in terms of the radial component of the random walk representing the polymer
chain.

We have to emphasize that all consideration in this paper was restricted to polymer
models without excluded volume in 3D. There is no doubt that self-avoidance, as well
as hydrodynamic interactions will significantly affect the specific scaling
relations obtained here. The account of excluded volume effects is, to some extent,
simpler than that of hydrodynamic interactions. On the basis of scaling approach we
can predict the pulling time, $\tau(N)\sim N^{1+\nu}/f$, for the polymer
characterized by the exponent $\nu$ in the relation $R\sim N^{\nu}$. Our
consideration is also restricted in the sense that we did not consider the role of
the polymer part already `swallowed' by the potential well.  Although this might
affect the result under some conditions, our aim was to elucidate the mechanism of
sequential straightening of folds, for that purpose the absorbed tail is irrelevant.
Similarly, in reality polymer is usually pulled into a hole in a wall, such as
membrane, and we have neglected the (presumably logarithmic) factors associated with
the excluded half-space.

However, as we said in the beginning, the purpose of the model is to facilitate
methods and intuition. In this sense, we think that the consideration of ideal
polymer in this letter was fruitful, because the mechanism of sequential
straightening of folds obviously applies to a number of real physical situations,
such as, e.g., DNA translocation through the membrane pore driven by the difference
in chemical potentials.

AYG gratefully acknowledges warm hospitality he felt throughout his stay at the
LPTMS where this work was done.  This work is partially supported by the grant
ACI-NIM-2004-243 "Nouvelles Interfaces des Math\'ematiques" (France).


\begin{thebibliography}{99}

\bibitem{homopolymer} K. Binder, in {\it Phase Transitions and Critical Phenomena},
vol. 10, ed. C. Domb and J. Lebowitz (New York: Academic, 1983)

\bibitem{Eisenregler}  {\it Polymers Near Surfaces}, (Singapore: WSPC, 1993)

\bibitem{gr_sh} A.Yu. Grosberg, E.I. Shakhnovich, Sov. Phys.–JETP {\bf 64}, 1284 (1986)

\bibitem{nech_na} A.Naidenov, S. Nechaev,  J. Phys. A: Math. Gen. {\bf 34}, 5625
(2001)

\bibitem{madr} N. Madras, S.G. Whittington, J. Phys. A: Math. Gen. {\bf 35}, L427
(2002)

\bibitem{step} S. Stepanow, A.L. Chudnovskiy,  J. Phys. A: Math. Gen. {\bf
35}, 4229 (2002)

\bibitem{forg} G. Forgas et al, in {\it Phase Transitions and Critical Phenomena}
eds. C. Domb and J Lebowitz (New York: Academic, 1991)

\bibitem{vannim} J. Vannimenus, B. Derrida,  J. Stat. Phys. {\bf 105}, (2001) 1

\bibitem{alex} K.S. Alexander, V. Sidoravicius,  {\tt eprint arXiv:math/0501028}

\bibitem{DeGennes1}  P.G. de Gennes J. de Physique Lett. {\bf 46}, L639 (1985)

\bibitem{DeGennes2}  A.Buguin, F.Brochard-Wyart, P.G.de Gennes, CR Acad. Sci.
Paris, {\bf 322 IIb}, 741 (1996)

\bibitem{klushin} L. L. Klushin, J. Chem. Phys. {\bf 108}, 7917 (1998).

\bibitem{obukhov}  C. F. Abrams, N. Lee, S. Obukhov, Europhys. Lett., {\bf 59}, 391
(2002)

\bibitem{gold}  A. Halperin, P. M. Goldbart, Phys. Rev. E {\bf 61}, 565 (2000)

\bibitem{Kasianowicz} J.J. Kasianowicz, E. Brandin, D. Branton, and D.W. Deamer,
Proc. Natl. Acad. Sci. U.S.A. {\bf 93}, 13770 (1996).




\bibitem{Meller3} A. Meller, L. Nivon, and D. Branton, Phys. Rev. Lett. {\bf 86} 3435 (2001).


\bibitem{Storm} A.J. Storm, C. Storm, J. Chen, H. Zandbergen, J.F. Joanny, and C.
Dekker, Nano Lett. {\bf 5}, 1193 (2005).

\bibitem{kar1}  J. Chuang, Y. Kantor, M. Kardar, Phys. Rev. E {\bf 65}, 011802 (2002)

\bibitem{kar2} Y. Kantor, M. Kardar, Phys. Rev. E {\bf 69}, 021806 (2004)

\bibitem{Pincus_blobs} P. Pincus, Macromolecules, {\bf 9}, 386 (1976)

\bibitem{sausage} M. Donsker, S.R.S. Varadhan, Comm. Pure Appl. Math, {\bf 28}, 525 (1975)

\bibitem{bessel1} J.D. Pitman, M. Yor, "Bessel process and infinetely divisible
laws, in {\it Stochastic integrals}, ed. D. Williams, {\it Lecture Notes in
Mathematics}, vol. 851 (Berlin: Springer, 1981)

\bibitem{bessel2} A.N. Borodin, P. Salminen, {\it Handbook of Brownian motion -- Facts
and Formulae}, 2nd ed. (Basel: Birkh\"auser, 2002)

\bibitem{verlet} W.C. Swope, H.C. Andersen, P.H.Berens, K.R. Wilson, J. Chem. Phys.,
{\bf 76}, 673 (1982)

\bibitem{KCh} J.P. Kemp, Z.Yu. Chen, Phys. Rev. E {\bf 60}, 2994 (1999)

\end{thebibliography}
\end{document}